\documentclass[prl,amsmath,showpacs,twocolumn]{revtex4}

\usepackage{graphicx,dcolumn,epsfig}

\begin{document}

\title{Suppression of modes in the random phase approximation}
\author{F. D\"onau}
\affiliation{Institut f\"ur Kern- und Hadronenphysik, 
             Forschungszentrum Rossendorf, 01314 Dresden, Germany}

\date{\today}

\begin{abstract}
A general but simple method is proposed to 
eliminate the quantum fluctuations generated by selected one-body 
operators in the excitation spectrum of  a discrete RPA Hamiltonian.
This method provides an outstanding tool for the removal of the 
contaminating spurious effects originated from symmetry violations. 
It can be also applied as a mode filter for analysising 
RPA response functions. 
\end{abstract}


\maketitle

\affiliation{Institut f\"ur Kern- und Hadronenphysik,  Forschungszentrum
Rossendorf, 01314 Dresden, Germany}

\begin{figure}
\epsfig{file=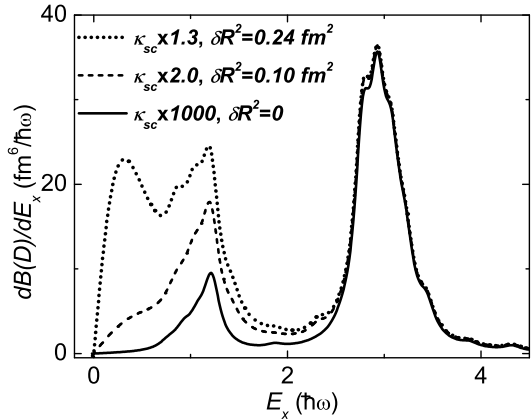,width=8.3cm,angle=0}
\caption{\label{fig:isoscalar} Illustration of the suppression of the
spurious c.m. fluctuations in the RPA response function of the
isoscalar dipole operator $D=r^3 Y_{1m}$ for a nuclear system with 
equal proton and neutron
numbers $N_\pi =N_\nu = 20$.
The RPA Hamiltonian consists of a spherical
Nilsson potential and a residual interaction of dipole plus octupole
type. The selfconsistent value $\kappa_{sc}$ \cite{BM}
of the isoscalar dipole strength is 
multiplied by an increasing factor in order  to demonstrate the disappearance of 
fluctuations {\boldmath $\delta R^2$} of the c.m. coordinate {\boldmath $R$}.
 In the solid curve 
the fluctuations of both the c.m. coordinate {\boldmath $R$} and 
the c.m. momentum {\boldmath $P$} are damped out.
}
\end{figure}

The Random Phase Approximation (RPA) \cite{Rowe, Ring} is a powerful 
standard approach  
to calculate microscopically the variety of vibrational excitations and 
giant resonances in nuclei and in other finite fermion systems like  
quantum dots or metal clusters \cite{RelRPA, CrankRPA, quantumdots}.
Based upon a stable equilibrium configuration of independent 
quasiparticles in a mean field (MF) potential,  
the RPA is accounting for the  
quantized small amplitude oscillations 
(named below as fluctuations)
about the MF equilibrium 
point by allowing Bose-like two-quasiparticle (2qp) excitations
driven by the residual interaction. 
There is rich diversity of phonon excitation modes produced 
in this way which shows up as complex resonance pattern 
in calculated RPA response functions.\\
 In this letter we present a new method which enables a forced 
damping of selected degrees of 
freedom. On the one side this method provides us  a 
perfect tool for the elimination of spurious modes as e.g.
the unwanted center of mass (c.m.) motion as examplified in Fig.~1 
or the fluctuations of the angular momentum  (a.m.) and the particle 
number as demonstrated in Fig.~2 which anyway
contaminate the calculated response functions.
On the other side the method can be used as a more 
general instrument for probing system properties by 
switching off the fluctuations of selected 
excitation modes for a spectral analysis
as demonstrated in Fig.~2 for a single
mode.\\
\begin{figure}
\epsfig{file=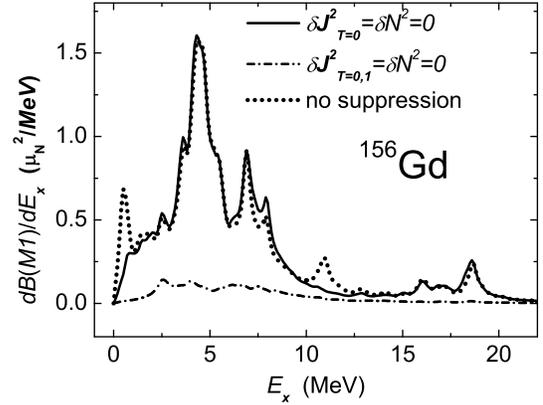,width=8.3cm,angle=0}
\caption{\label{fig:Gd156}Mode suppression effects illustrated 
in a schematic RPA calculation of the M1 response function for nucleus $^{156}$Gd.
The RPA Hamiltonian consists of a rotating axially deformed 
Nilsson plus pairing potential
and a residual interaction of spin plus quadrupole type. 
Solid curve: Fluctuations of both 
the isoscalar $(T=0)$ a.m. {\boldmath $J_\pi + J_\nu $} 
and of the particle numbers
$N_\pi,N_\nu$ suppressed. Dotted curve: without any suppression. 
Dashed dotted curve:
Like the solid curve but additionally the fluctuations of the 
isovector $(T=1)$ a.m. part  {\boldmath $J_\pi - J_\nu $}
 suppressed.
Note, that in the latter curve the dramatic effect might be not realistic
because of the simple interaction. 
}
\end{figure}
The underlying idea of our method is quite simple.
Let us consider a one-body operator $F$ which 
by definition incorporates
a specific mode of 2qp excitation. Then,  
by adding the oscillator-like restoring force term $\kappa F^2$ to the 
considered RPA Hamiltonian, one can shift arbitrarily the 
oscillator frequency 
by tuning the strength parameter $\kappa$. In the limit 
 $\kappa \rightarrow\infty$  the restoring force gets
so large  that the RPA fluctuations of the 
operator $F$ become completely suppressed, i.e. the operator $F$ is frozen
at its constant MF value. In this spirit the excitation of the 
mode $F$ gets eliminated from the RPA excitation spectrum.
Hence, by choosing appropriate 
force terms the corresponding excitation modes 
can be systematically filtered out (cf. Fig.~2). \\ 
Now the essential steps of our derivation will be sketched using  
well known relations \cite{Rowe,Ring, Blaizot}.\\
Let us consider a whole group of one-body operators  
$P{(r)}$ and $Q{(r)}$  selected for the removal of the associated  
modes (denoted below as $PQ$ modes). 
Examples for such operators are the c.m. 
coordinates {\boldmath $R$}, the particle number operator $N$ or 
the multipole operators $Q_{lm}=r^l Y_{lm}$. The $PQ$ operators
are given by
\begin{equation}
 P{(r)}=i\sum_i{p^{(r)}_i}(b^+_i-b_i)\,,\,\,\,Q{(r)}=\sum_i{q^{(r)}_i}(b^+_i+b_i)
\end{equation}
where we restrict ourselves without loss of generality to hermitean coordinate-like and 
momentum-like operators with any real coefficients $p^{(r)}$ and $q^{(r)}$.
In eqs.(1), $b_i$ and $b^+_i$ denote the basic 2qp excitation operators  
treated in RPA as bosons obeying approximately
the quasi boson approximation (QBA) $[b_i,b^+_j]\approx\delta _{ij}$
\cite{Rowe,Ring}. 
In practice, the explicit bosonic form, eq.(1)  
  follows from the original fermion representation of the $P$s or $Q$s by a 
qp transformation and using subsequently the QBA.
Possible non-zero expectation values of the quasiparticle operator
like the average particle number can be omitted since they are irrelevant 
constants. \\
Let be  $H(b,b^+)$ the bosonic Hamiltonian of the system under study 
derived analogously by applying the QBA. As mentioned above
the key of our removal procedure is to add an oscillator-like
force term to $H(b,b^+)$  in which the sum  of the squared $PQ$ operators
is collected:
\begin{equation}
    H'(\kappa ) = H(b,b^+)+\kappa \sum_r[{P{(r)}^2 + Q{(r)}^2}]
\end{equation}
The RPA excitation modes $s$ and their energies $E'_s$ 
are constructed by solving the equations of motion  as usual 
\cite{Rowe} but here for the supplemented Hamiltonian $H'$:
 \begin{eqnarray}
[H'(\kappa ),O^+(s)]=E'_s O^+(s),\\\nonumber
O^+(s)=\sum_i{[\,X_i(s) b^+_i-Y_i(s) b_i\,]}.
\end{eqnarray} 
The operators $O^+(s)$ denote the familiar 
excitation operators of the RPA eigenmodes and
the amplitudes $(X,Y)$ are built from the eigenvectors which 
diagonalize the well-known RPA matrix to $H'(\kappa )$.
It is irrelevant for our derivations 
that in general not all modes $s$ do
have normalizeable amplitudes (cf.\cite{Rowe}).\\
The tunable $PQ$ force term in the Hamiltonian $H'$, eq.(2) 
serves for shifting
the selected modes away from the energy region of the usual 
RPA excitations
in order to achieve the intended mode suppression. 
In this spirit our method 
extends the known idea of pushing out unwanted excitation 
modes from the physically interesting spectrum
as it was proposed, e.g., for the elimination of the c.m. motion  
in order to restore the Galilean invariance \cite{Thouless-spurious}.
Similar requirements were formulated 
with respect to the rotational spurious motion \cite{Marshalek}.
However, these options have not been pursued consequently 
in the past.\\
The common strength parameter $\kappa$ in eq.(2) is used as a 
scale variable to achieve
the crucial spectral shift of the $PQ$ modes. 
The scaled Hamiltonian $H'' $ given by 
\begin{equation}
                H''(\kappa )= \frac{H'(\kappa )}{\kappa}  =
                   \frac{ H(b,b^+)}{\kappa} 
                     +\sum_r[{P{(r)}^2 + Q{(r)}^2}]
\end{equation}
has obviously the same solutions $O^+(s)$ as $H'$ to the 
scaled eigenvalues $E''_s=E'_s/\kappa $.
In the limit $\kappa \rightarrow \infty$ the term  $H(b,b^+)/\kappa$ vanishes asymptotically 
and the spectrum of $H''$ is determined alone by the remaining supplementary term.
(Note, that this statement and its consequences below would be the same for the limit 
$\kappa \rightarrow -\infty$.)\\
The RPA eigenvalues $E''_s$ can be either exactly zero or have any real or 
complex values.
For a sufficiently large value $\kappa =\kappa_{max} < \infty$ the part of the eigenvalue
spectrum related to the scaled term $H(b,b^+)/\kappa$ will come close to zero 
whereas the dominating non-scaled supplementary term $\sum_r[{P{(r)}^2 + Q{(r)}^2}]$ has 
possibly exact zero eigenvalues (see below) or other eigenvalues but those certainly 
converging to a finite distance to the zero point. Accordingly, the corresponding spectrum of the 
non-scaled Hamiltonian $H'$, eq.(2) decomposes after rescaling $E'_s=\kappa E''_s $
into three parts: (i) the exactly zero eigenvalues, (ii) eigenvalues shifted to infinity both 
 related to the
supplementary term in eq.(2) and (iii) the ''normal'' vibrational eigenvalue spectrum 
related mainly with the original part $H(b,b^+)$
to be cleaned up from selected predefined $PQ$ modes. Therefore, by including a supplementary term 
and scaling we succesfully separate in the asymptotic region 
$\kappa=\kappa_{max} <\infty$ the normal phonon spectrum from the part of the $PQ$ modes. \\
The resulting separation of the energy spectrum for large enough values $\kappa$
translates in a partitioning of the 
RPA solutions. We denote the asymptoptic zero and infinity energy solutions as $O(s=a)=\Omega_a$ 
and $O^+(s=a)=\Omega^+_a$
and the normal vibrational solutions belonging to the finite energies
as $O(s=n)=\omega_n$ 
and $O^+(s=n)=\omega^+_n$. Since by construction the sets $\Omega _a ,\Omega^+ _a$ and $\omega _n,
\omega^+_n$
, repectively,  belong
to different energy regions these modes are orthogonal, i.e., they satisfy the orthogonality relations 
\cite{Rowe}:
\begin{equation}
[\Omega _a,\omega ^+_{n}]=[\Omega^+ _a,\omega ^+_{n}]=[\Omega _a,\omega_{n}]=0. 
\end{equation}
Defining $|0\rangle $ to be the normalized vacuum state to the normal solutions $\omega^+(n)$,  
the vibrational states $|n\rangle$ are created by the phonon excitation 
\begin{equation}
|n\rangle = \omega^+_n|0\rangle .
\end{equation}
Hence by using eq.(5) one obtains  
\begin{equation}
\langle 0 |\Omega^+ _a |n\rangle = \langle 0 |\Omega _a |n\rangle = 0  
\end{equation}
for all $a$ and $n$, respectively.\\
 We recall that in the asymptotic region $\kappa\rightarrow\infty$
 the  set $\Omega _a,\Omega^+ _a  $ can be considered as
 a RPA solution to the scaled  Hamiltonian $H''= \sum_r[{P{(r)}^2 + Q(r)^2}]$, eq.(4) 
such that one can construct these asymptotic solutions of $[H'',\Omega^+ _a]=E_a\Omega^+ _a$
directly in terms of the $PQ$ operators  
\begin{eqnarray}
\Omega^+_a = \sum_r [\tilde X_r(a)P(r)+\tilde Y_{r}(a)Q(r)]
\end{eqnarray}
instead of solving the full equations of motion (2).
Relying on the existence of a closure relation  \cite{Rowe} 
the above equation can be considered as formally inverted to get the expansions
\begin{eqnarray}
P(r)=\sum_a\tilde x_r(a)(\Omega^+_a-\Omega_a),\nonumber\\
Q(r)=\sum_a\tilde y_r(a)(\Omega^+_a+\Omega_a).
\end{eqnarray}
Hence, by using eqs.(7,9) we arrive at the crucial relations
\begin{equation}
\langle 0 |P(r)|n\rangle = \langle 0|Q(r)|n\rangle = 0 
\end{equation}
in the asymptotic region $\kappa \rightarrow\infty$ .
These relations express that all vibrational transition amplitudes 
for any selected $PQ$ operator vanish, i.e. all those 
operators cannot create vibrations from the phonon vacuum.
Thus, it is proven that the inclusion of a supplementary 
interaction term into the boson Hamiltonian enables one to eliminate the
fluctuations caused by all the operators $P(r)$ and $Q(r)$ from all the
normal solutions $\omega ^+_{n}$. In other words, the response function 
of any selected $PQ$ operator is identically zero in the 
physical spectral region.\\
Defining the sum rule $S_F$ of a linear boson operator $F$ 
(i.e. $F$ is a 2qp excitation operator) as 
$S_F=\sum_{n}|\langle 0|F| n\rangle|^2= \langle 0| F^2 |0\rangle
\equiv \delta F^2$ 
then eq.(10) can be written in compact form as   
\begin{equation}
\delta  P^2= \delta Q^2=0,  
\end{equation}
i.e. the complete disappearance of vacuum fluctuations in the asymptotic region 
$\kappa \rightarrow \infty$ for all $P$ and $Q$.
The above result (11) can be interpreted as a total blocking of RPA fluctuations
such that the momenta $P$ and coordinates $Q$ get fixed.\\
For illustration we present two simple but typical examples for the supplementary Hamiltonian (2).
The first example is the forced damping of the fluctuation for a single operator $P $ (or $Q$). 
Then, in the asymptotic region holds simply $H''=P^2$. One immediately realizes that the 
corresponding eigenmode is the ''zero vector''
$\Omega=\Omega^+=P$ with the eigenvalues $E''=E=0$ since $[H'',P]=0$. 
  The same applies also to a series of $P-$operators (or analogously
$Q-$operators) since $[P(r),P(r')]=[Q(r),Q(r')]=0$ are mutually commuting. One
gets a series of zero eigenvectors identical with the 
$P$s or alternatively $Q$s, respectively.\\
Choosing symmetry operators like the c.m. momentum, the a.m. operator 
or the particle number, then this mode suppression accomplishes the successful restoration 
of the corresponding symmetries on RPA level. 
For instance, when a statical pair field for protons and neutrons
are included in the MF,  
the fluctuations of the proton and neutron particle numbers can be removed by
using the supplementary term $(N_\pi ^2+N_\nu ^2$). \\
The second example concerns a conjugated operator pair  $(P,Q)$. 
Defining $\tilde Q=-Q/(2\sum_r p_r q_r)$
yields the commutator $[P,\tilde Q]=-i$ which leads to an oscillator-like
supplementary Hamiltonian $H''= 1/2(P^2+\tilde Q^2) $. 
Hence, the asymptotic eigensolution to $[H'',\Omega^+]=E''\Omega^+$ is explicitly given 
by $E''=1$ and
 $\Omega^+=1/\sqrt{2}(\tilde Q+iP)$ which yields   
$P=(\Omega^+ -\Omega)/\sqrt{2}$ and $Q=(\Omega^+ +\Omega)/\sqrt{2}$ and 
$E=\kappa E''$  indeed shifted to infinity for 
$\kappa\rightarrow \infty$. Taking e.g. for  the operator pair   
the c.m. momentum and coordinates then eqs.(11) realize the total removal of the spurious
c.m. fluctuations i.e. the c.m. coordinate is fixed in space and the system gets held at rest
(cf. Fig.~1).\\
Now we consider the possible numerical problems encountered in the RPA calculations 
when performing the above described
mode removal in practice. Firstly, it should be mentioned that the mode suppression (11)
by an asymptotic supplementary term  is working for any finite dimensional RPA Hamiltonian 
independently 
of whether the inherent mean field part is selfconsistently calculated or not. Further, it 
does not matter 
whether the RPA interaction consists of a sum of factorized terms like the well known 
multipole expansion or it is non-factorized like the nuclear Skyrme interaction. The actual form 
of the interaction
only determines whether the eigenvalue RPA matrix equation can be performed by the more 
convenient response function method 
\cite{separ-rpa} or by a more involved matrix diagonalization.\\ 
With respect to the actual performance of the transition $\kappa\rightarrow\kappa_{max} <\infty$
the critical question arises whether numerically  stable results can be 
expected or not.
The answer is clearly positive for a boson Hamiltonian $H(b,b^+)$ with a factorized interaction part.
 In this case all strength parameters enter inversely in 
the response  function matrix (cf. \cite{separ-rpa}). Hence,  
 the asymptotic limit of $1/\kappa$  gives a well defined zero 
which is causing no problem at all. 
The same should be true also
for a non-factorized interaction noting the latter can be expanded into a factorized series 
\cite{separ-rpa}. In fact, we studied the convergence properties of a straightforward RPA matrix 
diagonalization for  random RPA matrices up to the 
dimension $n=1000$ supplemented with a series of non-commuting 
random $PQ$ terms. It turned out that the stepwise variation of up to the considered value
$\kappa_{max}=10^4$ leads to a rapide convergence of the vanishing fluctuations (11) 
together with converging RPA eigenvalues $E'_s$ 
which points to a good numerical performance. 
How large the value $\kappa_{max}$ should be actually taken can be easily 
controlled by checking the 
descending size of the zero point fluctuations $\delta  P^2$ 
and $\delta Q^2$. \\
We are now turning  to  possible physical applications and improvements seen 
so far for the nuclear structure 
investigations done with RPA. \\
The mode suppression should be included regularly for the removal of spurious modes 
connected with violations of conservation rules as for instance the Galilean invariance, 
a.m. conservation due to a deformed and possibly rotating mean field, 
particle number conservation broken by a static pair potential
etc. to avoid the corresponding fluctuations.
We mention that our method is exact as compared to the approximate symmetry restoration  
\cite{Pjatov}
which in general can not satisfy the condition (11) for a complete suppression of 
the spurious motion. \\
There are immediately some useful consequences of our method. 
For instance, the effective charges $e_\pi $  and $e_\nu$ 
of the the electric dipole operator 
usually introduced by a transformation to the c.m. system \cite{Ring} become obsolete
 as well as the c.m. correction  
$Q_{3m}=r^3 Y_{3m} -\eta  rY_{1m}$  
of the octupole operator discussed in \cite{hamamoto-spurious} is not needed 
(cf. Fig.~2)
when the c.m. mode {\boldmath $R$} is suppressed.\\ 
For a rotating deformed system 
the mode suppression enables one to define cleanly
a body fixed or alternatively an a.m. fixed reference system. 
The body fixed system is singled out by the requirement
$\delta  Q^{2}_{2m=\pm 1}=\delta J_0^2=0$ which accomplishes 
that both the 
principal axes system \cite{Ring} 
as well as the length $|$\boldmath$J$$|$ of the a.m.
are conserved in the RPA. 
Alternatively, the damping out of the fluctuations of the a.m. vector through
the condition $\delta$\boldmath$J$$^2$=0 determines the lab reference 
by fixing length and direction of the a.m. vector \boldmath$J$ 
 but still leaving freedom for possible
shape and orientation fluctuations of the rotating deformed nucleus.
The latter option could be an important ingredient for improving the RPA description of  
wobbling excitations 
found in triaxially superdeformed rotating nuclei \cite{wobbling}.\\
Another important new aspect is that a simplified version of the
RPA can be utilized where the MF part is adopted from the extremely 
successful semimicroscopic models that apply the Strutinsky shell 
correction method \cite{Ring}.  
The bosonic interaction part might then be added as a 
multipole-multipole interaction with adjustable 
strength parameters. The suppression  of spurious modes works also for this
case and when properly treated this simplified RPA satisfies the 
stability criterion \cite{Thouless-spurious}.\\
The mode suppression enables one to perform also a RPA calculation
based on a 2qp excited state instead on the MF ground state.
Usually, this interesting new possibility is excluded since excited MF configurations 
violate the RPA stability criterion
\cite{Thouless-spurious}. However, the stability can be recovered
by a force term that is blocking explicitly the deexcitation to the ground state 
configuration.\\
Finally, we like to emphasize that the proposed mode 
suppression causes almost no additional 
effort compared to a RPA calculation without supplementary terms.\\
In summary, the restoration of symmetry violations is one field of applications of the proposed 
mode suppression. The possible application as a spectral filter   
offers as another exciting new field. Since the full RPA spectrum implies in general a 
huge number of states the mode suppression provides us a new tool for a better 
identification of distingished modes. 
This seems to be mandatory for a more reliable interpretation of 
fine structures in the response functions in comparison with measured data. \\ 
The author acknowledges thankfully worthwile discussions with Dr. R. G. Nazmitdinov.

\end{document}